# Astro2020 Science White Paper

# Developing a vision for exoplanetary transit spectroscopy: a shared window on the analysis of planetary atmospheres and of stellar magnetic structure

**Thematic Areas:** ☒ Planetary    ☒ Stars and Stellar

**Principal Author:**
Name: Adam Kowalski
Institution: University of Colorado/National Solar Observatory/Laboratory for Atmospheric and Space Sciences
Email: adam.f.kowalski@colorado.edu
Phone: 419-704-7509

**Co-authors:**
Karel Schrijver
Valentin Martinez Pillet (National Solar Observatory)
Serena Criscuoli (National Solar Observatory)

**Abstract:**
We describe how the accurate characterization of exoplanetary atmospheres in the ELT and JWST era will inevitably require taking into consideration of the stellar inhomogeneities caused by convection and magnetic fields. The existing evidence that demonstrates the mixture of stellar and planetary signatures in observed transiting spectra is presented. Finally, we discuss how to disentangle these two components through a multipronged approach that includes new solar reference spectra, improved MHD modeling, and synergistic collaborations between the communities involved, from solar to stellar and exoplanet astronomers.

# Transit spectroscopy of exoplanets as a window to stellar and exoplanet characterization

There are about 3,000 confirmed transiting exoplanets (as reported by the NASA exoplanet archive) with radii ranging from at least double the size of Jupiter —itself at 11 Earth radii— to half the size of Earth. These, and many more expected to be discovered by, e.g., TESS, are all candidates for atmospheric studies using their star's light that travels through the transparent layers of the exoplanetary atmospheres (Figure 1). Spectroscopic analyses are already applied to some, and tested for others, to learn about exoplanetary atmospheric sizes and densities (up into their exospheres), to establish their chemical compositions, and even to constrain cloud covers, hazes, exoplanetary rotation, and large-scale exo-atmospheric winds (Pinhas et al. 2018).

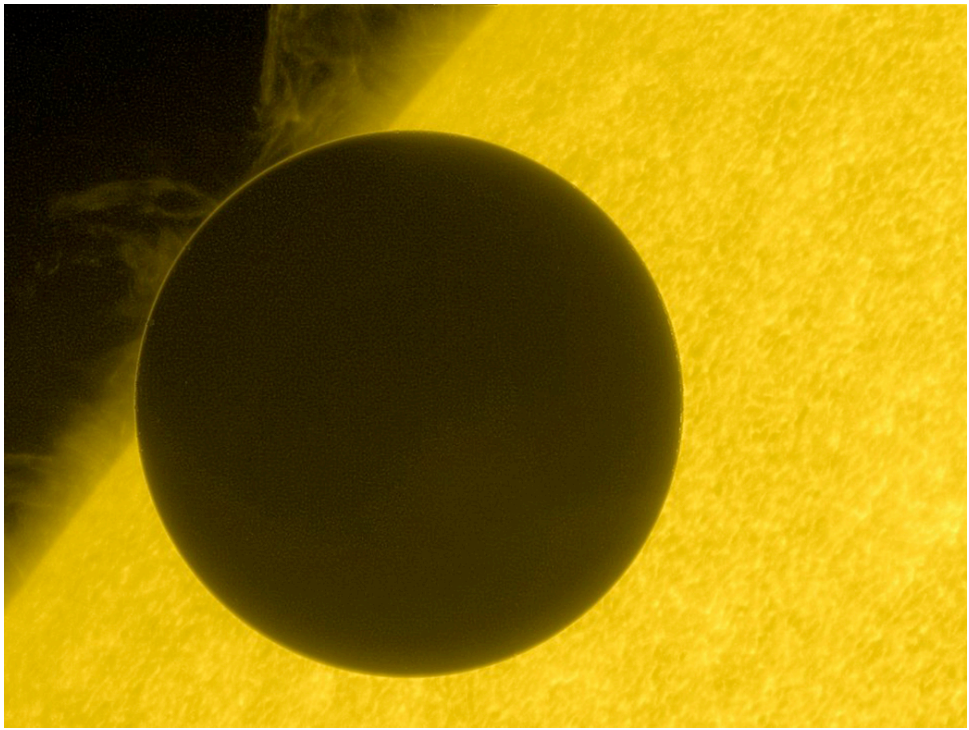

*Figure 1: Venus's transit observed against inhomogeneities resulting from convective* motions at the solar surface. Note the Venus atmospheric rim produced by scattered solar light. Image courtesy of Hinode (JAXA/NASA/ESA).

Transiting spectroscopy is a powerful, new technique to probe stellar surface inhomogeneities, thus effectively achieving an indirect means of spatially resolving other stars. Outside of significant flare activity, stars with convective envelopes exhibit a broad range of magnetically active coronae, from $\log\frac{L_X}{L_{bol}} \sim -7$ for the Sun and the solar-type star alpha Centauri A (Ayres 2015) to $\log\frac{L_X}{L_{bol}} \sim -3$ for the most active stars (Pizzolato et al. 2003, Wright et al. 2016). These dramatic coronal differences should manifest in similarly conspicuous photospheric and chromospheric emission signatures, such as spots, plage, bright points, and faculae with distinct spectral signatures as they do on the Sun.

Until recently, the common assumption was that stellar surfaces were unblemished and that only an approximate center-to-limb gradient in their spectral signature had to be considered during transits. But with advancing signal-to-noise (S/N) ratios as larger telescopes (extremely large telescopes; ELTs, and the James Webb Space Telescope; JWST) and dedicated instrumentation are being brought to bear, it is clear that this is no longer adequate if exoplanetary atmospheres are to be analyzed (see, e.g., Aronson 2019). Transits are being used to test our understanding of stellar atmospheric dynamics (granulation patterns) and the resulting stratification, which together set the center-to-limb behavior (Dravins et al. 2018). And it is being recognized that occulted areas during an exoplanetary transit can cover quiescent photospheric regions, the stellar equivalent of active regions, starspots, and large starspot clusters (Morris et al. 2017, Rackham et al. 2018).

On the Sun, spots and quite sun regions show very different spectral signatures (see the atlas by Wallace et al. 1994) for two reasons. First, different thermal stratifications (by several thousands of Kelvins) give rise to different atomic and molecular populations with drastically different spectral signatures. Second, the presence of a magnetic field broadens, or even splits, spectral lines into multiple components due to the Zeeman effect separated by fractions of an Angstrom in the case of the strongest kG fields, which are also detected in other magnetically active stars (Johns-Krull & Valenti 1996, Reiners 2012). Thus, over the next decade transit spectroscopic observations open two avenues of scientific discovery side by side: the properties of the exoplanets' atmospheres and the characterization of stellar inhomogeneities, with the Sun and the Solar System inner planets' transits as a Rosetta stone ready to help de-cypher the distant transit signals.

The planets function as 'occulting disks' projected on stellar surfaces during their transits, while exoplanet atmospheric details can be disentangled from the 'contamination' in the transit spectra that result from unexplored signals from the transited stars. However, two aspects will undoubtedly complicate transit spectroscopic interpretations going forward. On the one hand, transits offer us unprecedented access to the structure of stellar magnetic fields: whereas techniques like (Zeeman) Doppler Imaging and rotational modulation analyses offer, at their very best, a resolution of 100,000 km on a Sun-like star —provided that it rotates significantly faster than our Sun— a transiting occulting disk offers a resolution limited only by S/N and by the evolution of the source. On the other hand, exoplanetary atmospheric studies have access presently only to a limited sample of spectra from solar surface features. We simply do not know what such features look like on stars of different spectral types and rotation rates, or how such features evolve in time (from formation to decay on time scales of days to weeks, to migratory patterns under the transit paths in the equivalent of the Sun's butterfly pattern over the sunspot cycle). In fact, Zeeman broadening measurements of the most active (dMe) stars indicate that ~50% of the surface is covered with ~4 kG fields ($\sum \{Bf\}$=3.9 kG where f is the filling factor; Johns-Krull & Valenti 1996, see also review in Reiners 2012), which is a regime of magnetism that is not seen on the Sun. Thus, the question arises as to what types of surface inhomogeneities are present in other stars and how do their atmospheric structures and the ensuing spectra inform us about what we observe in transit spectroscopy. Sophisticated MHD modeling of the atmospheric stratification for different chemical compositions and stellar parameters (Trampedach et al. 2013; Beeck et al. 2015) will certainly help disentangle stellar and planetary spectroscopic signatures.

Decoupling stellar and exoplanetary contributions to observed transit spectra will not be easy, but it provides unexplored riches in information. Experiments are already underway (Pinhas et al. 2018; Dravins et al. 2018; Rackham et al. 2018; Sanchis-Ojeda et al. 2013; Cauley et al. 2018; Giovanni et al. 2018; Beky et al. 2014; Dai et al. 2018). Studies can start with Jupiter-sized planets, which would occult 1% of the disk of a Sun-like star, transiting in a matter of hours (e.g., 2.3h for an equatorial crossing of the "hot Jupiter" WASP-18b at 0.05AU from its star, and 8.2h for the cooler Jupiter-sized planet COROT-9b at 0.41AU). An Earth-sized planet would at present be a substantial challenge with a coverage fraction of only 0.01%, but there are thousands of intermediate-sized planets between these sizes to study as larger and more sensitive instrumentation is developed.

**This development is essential to Astrophysics (understanding planetary formation and evolution, from atmospheres to orbits), to Heliophysics (understanding the Sun's past, present, and future magnetic activity), to Planetary Physics (developing comparative atmospheric studies), and to the search for extraterrestrial life (by knowing how to separate abiotic from biotic spectral signatures). As a community, we need to explore the potential of discovery and to outline a roadmap of required investments in instrumentation and methods.**

As already discussed, the feasibility of transiting spectroscopy and photometry has been demonstrated for a number of stars with a range of magnetic activity levels, and the prospects will increasingly grow with the launch of the JWST. An analysis of transiting Kepler photometry of an active K4 dwarf HAT-P-11 has for the first time revealed a starspot distribution with active latitudes like the Sun (Morris et al. 2017). This result supports a magnetic dynamo mechanism that is similar to the Sun, and it validates flux transport models that employ solar-like injection patterns for simulating other stars (Schrijver 2001). The sizes of spots are also constrained by transiting photometry, revealing an average spot size similar to the Sun's at maximum but with a tail to the distribution extending to much larger sizes than observed on the Sun (Morris et al. 2017). The work by Pinhas et al. (2018) is also a pathfinder. They explored the role of stellar surface inhomogeneities in characterizing planetary atmosphere properties from transit spectra. By modeling the stellar "background", we obtain information about the atmospheres of the planets, such as their cloud and haze composition in addition to the pressure variation vs. height.  Pinhas et al. (2018) varied the surface coverages and temperatures of PHOENIX (radiative-equilibrium) models of hot faculae, photosphere, and cool spots to explore the variation in transit depth as a function of wavelength.  Thus, we obtain information about the stellar surface features and planetary atmospheres simultaneously.   Depending on the mix of stellar inhomogeneities, the transit depth differences are several tenths of a percent and are largest in the near-UV and blue.  The authors suggest the following:

*"While the effects of stellar heterogeneity can be significant at wavelengths as long as 2 µm (see Figure 4), they are most pronounced between 0.3 and 1 µm and thus high-impact observations should focus on this spectral range to probe the activity of stellar photospheres."*

Interestingly, they found that the most active star (according to its Ca II emission) in their sample has one of the most homogeneous surfaces, which could be a result of noisy transit data or simplistic modeling assumptions of the stellar background.

Ground-based solar physics can contribute to the transiting spectroscopy efforts by providing benchmark spectra for representative solar surface features, such as faculae, umbrae, penumbrae, granulation, and inter-granule lanes over a broad spectral range and viewing conditions.  Current solar models provide insufficient accuracy due to the large amounts of unknown opacity sources in the blue and NUV (Bruls et al. 1992, Fontenla et al. 2015, Criscuoli, 2019).  There are varying opinions concerning whether very high spatial resolution data (e.g., with the incoming Daniel K Inouye Solar Telescope) are needed for these benchmark spectra.  Other existing medium aperture solar telescopes can dedicate significant fractions of their time to obtain these reference spectra if adequately funded and instrumented. However, no current or future planned ground-based solar instruments have the full required capability for achieving both high spectral resolution and broad spectral coverage from the near-UV through the near-IR, as was possible in the 1980s with the Universal Spectrograph or the Fourier Transform Spectrometer at McMath Pierce Solar Telescope on Kitt Peak.

## Programmatic consideration

Progress in this field will require support to a multipronged approach that includes all of the following aspects:

- Establish ways in which the solar, stellar, and exoplanet communities effectively exchange information and address each other's needs by producing a roadmap to understand the impact of stellar inhomogeneities in transit spectroscopy.
- Dedicate at least one existing solar facility to obtain reference spectra of distinct regions on the Sun with different viewing angles, and over broad spectral range. This aspect will require new instrumentation.
- Test transit spectroscopy with Mercury and Venus examples.

- Support MHD modeling of different stellar types and generate reference synthetic spectra using careful NLTE line-profiles calculations.